\documentclass[jpcfk,twocolumn,groupeaddress,prb]{revtex4}
\usepackage{graphicx} 

\usepackage{amsmath}

\usepackage{amsfonts}
\usepackage[version=3]{mhchem}

\usepackage{dcolumn}

\usepackage{bm}

\newcommand{\dif}{\mathrm{d}}

\begin{document}

\title{Octa-coordination and the Hydrated Ba$^{2+}\text{(aq)}$ Ion }

\author{Mangesh I. Chaudhari}
\email{michaud@sandia.gov}
\affiliation{Center for Biological and Material Sciences, Sandia National Laboratories,  Albuquerque, NM, 87123}

\author{Marielle Soniat}
\email{msoniat@uno.edu}
\affiliation{Department of Chemistry, University of New Orleans, New Orleans, LA, 70148 }

\author{Susan B. Rempe}
\email{slrempe@sandia.gov}
\affiliation{Center for Biological and Material Sciences, Sandia National Laboratories,  Albuquerque, NM, 87185}

\date{\today}

\begin{abstract}

{The hydration structure of Ba$^{2+}$ ion is important for understanding blocking mechanisms in potassium ion channels. Here, we combine statistical mechanical theory, {\it ab initio} molecular dynamics simulations, and electronic structure methods to calculate the hydration free energy and local hydration structure of Ba$^{2+}\text{(aq)}$. 
The predicted hydration free energy (-304.10$\pm$1.38 kcal/mol) agrees with the 
experimental value (-302.56 kcal/mol) when the fully occupied and exclusive inner solvation shell is treated.
In the local environment defined by the inner and first shell of hydrating waters, Ba$^{2+}$ is directly coordinated by eight (8) waters.  
Octa-coordination resembles the structure of Ba$^{2+}$  and K$^+$ bound in potassium ion channels, but differs 
from the local hydration structure of K$^+\text{(aq)}$ determined earlier.}
\end{abstract}

\maketitle





\section{Introduction} 

Barium (Ba$^{2+}$) is about the same size as potassium (K$^+$) (radii within 0.02 \AA).  Both ions 
partition between water and  octa-coordinated
binding sites in potassium ion channels.\cite{Jiang:2000,lockless,ye,Jiang:2014,degroot}  
Thus, Ba$^{2+}$ can act as an analogue of K$^+$.
Octa-ligation by oxygens along the protein backbone of 
 K channels is widely believed to facilitate K$^+$ permeation by mimicking that ion's local
 hydration structure,\cite{Zhou:2001vo,Piasta:2011bu,Jiang:2014} 
even though experimental and theoretical studies report lower K$^+$ hydration numbers.\cite{Varma:2006} 
While  K$^+$ permeates rapidly, Ba$^{2+}$ instead sticks and blocks permeation of other ions.
That inhibitory behavior has been used since the 1970's to probe the mechanism of K channel function.
\cite{hagiwara,Armstrong:1980,
Eaton:1980,Armstrong:1982, Vergara:1983,miller:1987a,
miller:1987b,miller2,
miller1,isacoff,
Vergara:1999,alagem,ashcroft,moc:2005,moc:2008,lockless,Chatelain:2009,Piasta:2011bu}  
Yet, recent works still debate the blocking mechanism and structure of the blocking sites in 
various K channels.\cite{Kim:2011iv,rossi,roux}  To help clarify the debate, we analyze here the physical chemical properties of Ba$^{2+}$ in aqueous solution, the reference environment for ion channel block.  An unresolved question addressed is whether Ba$^{2+}$ hydration structure resembles K$^+{(\mathrm{aq})}$ or K channel  binding sites.


In a 1933 landmark theoretical paper on water and ionic solutions, Bernal and Fowler predicted an 
octa-coordinated Ba$^{2+}$ hydration structure.\cite{bernal}
Structural data to test that prediction is sparse. 
One reason is that Ba$^{2+}$ extensively absorbs X-rays, leading to unfavorable conditions
for structural studies.\cite{ohtaki}   
Consequently,  only a couple  
extended X-ray absorption fine structure (EXAFS) spectroscopy  experiments
have targeted Ba$^{2+}$ ion, producing a hydration number of eight (8).\cite{persson,dangelo}  
Similarly, theoretical analysis of Ba$^{2+}$ hydration using {\it ab initio} 
methods has been limited, partly due to the large number of electrons involved.\cite{rode:ba} 
One study combined an {\it ab initio} quantum mechanical approach with classical molecular mechanics simulations (QM/MM) and reported a hydration number of nine (9).\cite{rode:ba}
That structural result was not substantiated by a prediction of hydration free energy.

A statistical mechanical theory developed earlier permits computation of
 solvation free energy based on local structural results determined for
systems treated with {\it  ab initio} models.\cite{redbook,Beck:2006wp,Asthagiri:2010,Rogers:2011,Rogers}  
The coupling of structure with thermodynamic  predictions
provides an advantage for validating the results compared to more standard
studies of structure alone.  Also, structural data obtained by molecular simulation 
 contains information about the spatial distributions of each neighboring solvent molecule.  
Although a neighborship analysis is typically unresolvable for experimental data and 
seldom applied to simulation data, the results can help clarify how many solvent molecules define the
local hydration structure around an ion.

The free energy theory has been coupled with structural studies to obtain new insights about  
K$^+$,\cite{Rempe:K,Asthagiri:divalents,Varma:jacs} its monovalent analogue, rubidium (Rb$^+$),\cite{Sabo:2013gs} 
several other hydrated mono- and di-valent metals,\cite{Rempe:Li,Rempe:Na,Asthagiri:2003kl,Asthagiri:divalents,alam:li,Jiao:2011} 
and other solvation problems.\cite{Ashbaugh:Kr,sabo:h21,sabo:h2,clawson:h2,Varma:bj,varma:valino,Jiao:2011b,Jiao:2012,mic:kr}   Here, we take the same approach to increase our understanding of Ba$^{2+}{(\mathrm{aq})}$.
At the same time, we also investigate how parameter choices in the theoretical analysis affect the free energy predictions.

Our studies suggest that the best hydration free energy predictions result from analysis of the fully occupied
and exclusive inner solvation shell.   Contributions from both local and distant solvent molecules are important
for predicting the total hydration free energy.
Barium ion directly coordinates with $n$=8 waters in aqueous solution, 
which differs  from the  $n$=4 inner-shell coordination found earlier
for K$^+$.\cite{Rempe:K,Asthagiri:divalents,Varma:jacs}  In contrast to conventional
propositions, octa-coordinated ion binding sites in K channels appear to mimic the local hydration structure of Ba$^{2+}$, the blocking ion, not K$^+$, the permeant ion.

\section{Theory}

Quasi-chemical theory (QCT) \cite{redbook,Beck:2006wp,Asthagiri:2010,Rogers:2011,Rogers}  
 divides the \emph{excess} chemical potential of Ba$^{2+}$ hydration, $\mu^{(\mathrm{ex})}_{\mathrm{Ba^{2+}}}$, into three contributions, 

\begin{multline}
\mu^{\mathrm{(ex)}}_{\mathrm{Ba}^{2+}} = -kT\mathrm{ln}K^{(0)}_{n}\rho^n_{\mathrm{H_2O}}+kT\mathrm{ln}p_{\mathrm{Ba}^{2+}}(n) 
\\  +(\mu^{\mathrm{(ex)}}_{\mathrm{Ba(H_2O)}_{n}^{2+}}-n\mu^{\mathrm{(ex)}}_{\mathrm{H_2O}}).
\label{eq:1}\end{multline}
The first term represents an equilibrium ratio $K^{(0)}_{n}$ for Ba$^{2+}$-water association reactions (Eq. \ref{eq:2}) treated 
as in an ideal gas phase, hence the superscript $(0)$:  
\begin{equation} 
\mathrm{Ba}^{2+} + n \mathrm{H_2O} \rightleftharpoons \mathrm{Ba(H_2O)}_n^{2+}.\label{eq:2}
\end{equation}
The densities of water in solution, $\rho^n_{\mathrm{H_2O}}$, account for the availability of water ligands.  In the second term,
$p_{\mathrm{Ba^{2+}}}(n)$ is the probability of observing $n$ waters in an inner solvent shell of radius $\lambda$. This population fluctuation term will contribute
 zero if the inner shell is strictly defined by a single coordination number. 
The probability is readily evaluated from {\it ab initio} molecular dynamics (AIMD) simulations.  
The third term represents solvation of the Ba$^{2+}$-water inner-shell complex by the outer solvation
environment, and removal of the water ligands from the same environment. That combination,
 $\mu^{\mathrm{(ex)}}_{\mathrm{Ba(H_2O)_n^{2+}}}-n\mu^{\mathrm{(ex)}}_{\mathrm{H_2O}}$, balances the free energy for the ion-water association reaction.  The Boltzmann factor, $k$, and temperature, $T$, set the energy scale.

The hydration free energy is  independent of the $n$ and $\lambda$ parameters in QCT (Eq.~\ref{eq:1}).  
Nevertheless, some parameter choices are more convenient than others in practical applications that 
evaluate the free energy contributions approximately.  
One goal of this study is to determine how different 
choices for $n$ and $\lambda$ affect QCT 
predictions. 
For that analysis, we adopted the standard 
$\emph{no~split~shell}$ approach for separating inner- and outer-shell contributions 
to the hydration free energy (Eq.~\ref{eq:1}).
Accordingly, we defined the inner-shell solvent domain as the region containing the full subset of 
ligands directly coordinated with the ion,
as determined by  AIMD simulations. That definition excludes ligands that occupy both inner and outer
solvation shells. Then,  we compared hydration free energies computed with
seven $\lambda$ values within that inner-shell region  (2.9-3.5 in increments of 0.1~$\mathrm{\AA}$) and 
for coordination numbers that span the full range of possible inner-shell occupancy for Ba$^{2+}$.  
For comparison, we also considered for the first time a {\em multi-shell} occupancy consisting of the fully occupied inner shell
and one ligand in the outer solvation shell.

\section{Computational methods}

The principle challenge in predicting local hydration structure and association free energies for ions is to 
represent the broad range of molecular interactions involved in ion complexation reactions.
Previous studies suggest that treatment of
multi-body interactions is important for those predictions.\cite{whitfield,Varma:2010,bostick}   
Accordingly, we chose to model the hydrated Ba$^{2+}$ using density functional theory (DFT) since that
approach naturally accounts for interactions between pairs and larger groups of atoms.

  \begin{figure}
\begin{center}
\includegraphics[width=3.0in]{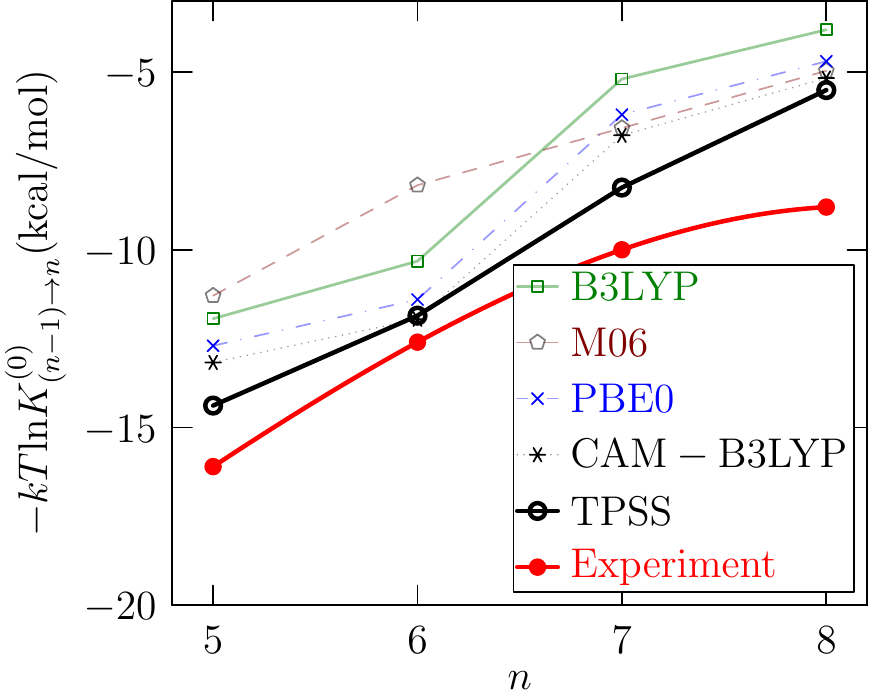}
\caption{ Free energies for sequential gas phase association reactions as a function of coordination number 
using several exchange-correlation density functionals.  
The reactions are: $\text{Ba}^{2+}\text{H$_2$O}_{n-1} + \text{H$_2$O}  \rightleftharpoons  \text{Ba}^{2+}\text{H$_2$O}_{n}$.
The red line represents experimental results from collision-induced mass spectrometry.\cite{Peschke1998}
The density functionals are (see inset): B3LYP, M06, PBE0, CAM-B3LYP, and TPSS.\cite{b3lyp,b3lyp2,m06,pbe0,cam,tpss}
Results were obtained with aug-cc-pvtz (O, H) or aug-cc-pvdz (on H for TPSS) basis sets and the MBW64 (Ba) effective core potential
and corresponding basis.
Of the functionals tested, the TPSS results come closest to the experimental results.}
\label{fig:pnx1}
\end{center}
\end{figure}

$Ab$ $initio$ molecular dynamics (AIMD) simulations on a single Ba$^{2+}$ solvated by 64 waters were performed using the
VASP AIMD simulation package.\cite{kresse} A cubic box of 12.417 $\mathrm{\AA}$ was used with periodic boundary conditions to mimic bulk liquid conditions. 
In this box volume, the water density matches the experimental density of liquid water at standard conditions. 
We utilized the PW91 generalized gradient approximation to the electron density,\cite{perdew} described the core-valence 
interactions using the  projector augmented-wave (PAW) method,\cite{blochl} expanded the valence electronic orbitals in 
plane waves with a high kinetic energy cut-off of 36.75 Ry (500 eV), used 10$^{-6}$ eV as the convergence criteria for the 
electronic structure self-consistent iterations, used a time step of 0.5~fs, and simulated the system in an NVE ensemble for 30~ps. During the simulation time, 
the average temperature was 326 $\pm$ 17 K.  That temperature helps avoid the over-structuring of water  observed in 
AIMD simulations of pure water at room temperature.\cite{rempe:water}
Prior to the production run, the system was equilibrated for $\approx$ 11 ps in an NVT ensemble with a constant temperature of 330 K.

Electronic structure calculations on single waters and clusters of Ba(H$_2$O)$_n^{2+}$ were 
performed using 
Gaussian 09\cite{g09} to estimate the equilibrium constants, $K^{(0)}_{n}$, for 
 inner-shell association reactions between Ba$^{2+}$ and water, as well as the outer-shell solvation contributions,
  $\mu^{\mathrm{(ex)}}_{\mathrm{Ba(H_2O)_n^{2+}}}-n\mu^{\mathrm{(ex)}}_{\mathrm{H_2O}}$. 
  Experimentally determined gas phase  free energies for sequential addition of waters to Ba$^{2+}$ provided data\cite{Peschke1998}
   for selecting the exchange-correlation functional and basis sets (FIG.~\ref{fig:pnx1}). 
  Based on comparison of several density functionals, we selected the TPSS\cite{tpss} exchange-correlation density functional, 
   with the aug-cc-pvtz (O) and aug-cc-pvdz (H) basis sets\cite{basis}   
   and the MWB46\cite{mwb46} (Ba) effective core potential and corresponding basis set. 
        Cluster conformations were exhaustively sampled to obtain optimized geometries and electronic energies.  
   Tight convergence criteria on the optimization (10$^{-5}$ a.u.)  and energy (10$^{-8}$ a.u.), along with an ultra-fine integration grid, facilitated
   the optimization procedure.
  Vibrational frequency analysis based on the normal modes\cite{Rempe:1998} were performed to obtain thermal corrections to the electronic energy.
  All vibrational frequencies were positive, confirming that optimized cluster configurations represent minimum-energy geometries.

  To evaluate the electrostatic contribution to the outer-shell solvation term of Eq. 1,
the radius of the barium atom was modified to match the chosen $\lambda$ values. 
The ion-water complexes were re-optimized in the presence of the environment,
treated here as a reaction field with a polarizable continuum model (PCM).\cite{Tomasi:2005tc}
 Default parameters were used for hydrogen and oxygen radii to create a solute cavity using a set of overlapping spheres. 
 The dielectric constant of the outer-shell environment was set to mimic water (78). Hydration free energies were calculated at 
 $T$=298 K and $p$=1 atm and subsequently adjusted 
 to account for the actual concentration of water ligands in liquid water, $\rho_{\mathrm{H_2O}}$=1 g cm$^{-3}$.  If this
 density is tracked as an adjustment to the ideal gas pressure, then it corresponds to a pressure factor of
 1354 atm.\cite{redbook,PaulGrabowski:2002cf} 

\begin{figure}[h*]
\centering
\includegraphics[width=3.0in]{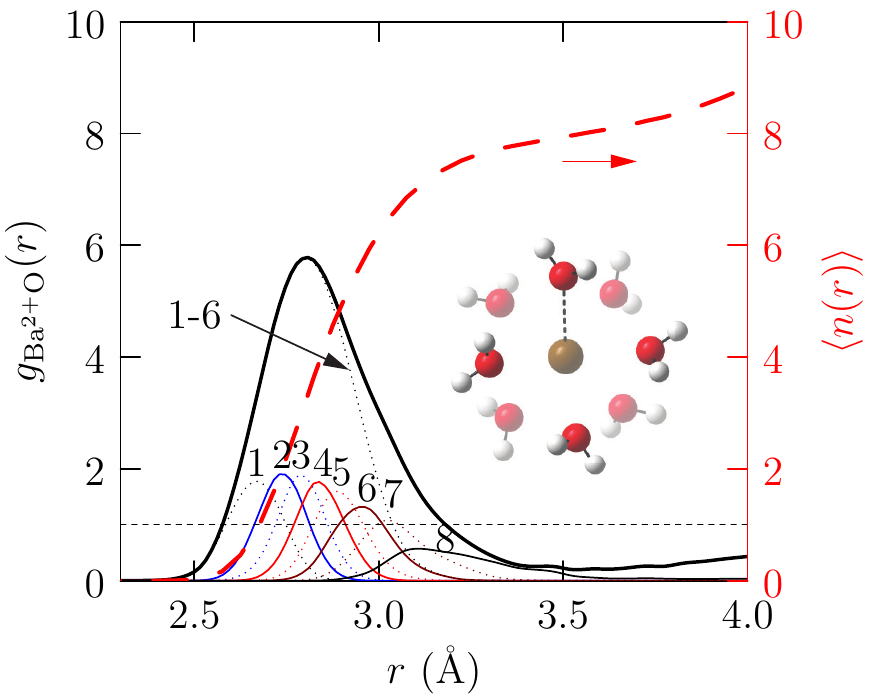}
\caption{Radial distribution function and near neighbor distributions for Ba$^{2+}$ ion in water using AIMD simulation results (30~ps). The picture illustrates
one observed conformation of Ba$^{2+}$ ion (brown) with $\left<n\right>$=$8$ water molecules (red oxygens, silver hydrogens) in the inner solvation shell. 
Four waters appear in front (bright colors) and four in the rear (light colors), creating a skewed cubic geometry. 
$\left<n(r)\right>$=$4\pi\rho_{\mathrm{O}}{\int_{0}^{r}}g_{\mathrm{Ba^{2+}O}}(r)r^2\dif{r}$ represents 
the running coordination number (red dashed line).  The inflection in $\left<n(r)\right>$ at $r$=3.5~{\AA} indicates that
eight near water neighbors stably occupy the  inner and first hydration shell of Ba$^{2+}\text{(aq)}$.}
\label{fig:gr}
\end{figure}

\section{Results and discussion}

The hydration structure determined by the Ba$^{2+}$-oxygen radial distribution function, $g(r)$,
 shows a distinct division between inner- and outer-shell water neighbors (FIG.~\ref{fig:gr}).   
Analysis of the near neighbor distributions reveals that
the first $n$=1-6 waters fill in the principal maximum at $r _\mathrm{max}\approx$ 2.8 $\mathrm{\AA}$.
The $n$=7$^{th}$ and $n$=8$^{th}$ near neighbors contribute to the first peak in a unimodal way, satisfying the {\it no~split~shell} rule
 and making the inner and first hydration shells identical.
 The $n$=9$^{th}$ neighbor occupies the second hydration shell.  
  An inflection point on the running coordination number $\left<{n(r)}\right>$ at the first minimum in the radial distribution function, $r_\text{min}\approx$ 3.5~{\AA}, confirms  
$\left<{{\bar{n}}}\right>=8$ as the average and most probable hydration number.     
EXAFS studies report the same  $\left<{{n}}\right>$ and $r_\mathrm{max}$ for aqueous 
 solutions of 0.8~M \cite{persson} and 0.1~M \cite{dangelo}  BaCl$_2$.

\begin{figure}[h*]
\begin{center}
\includegraphics[width=1.5in]{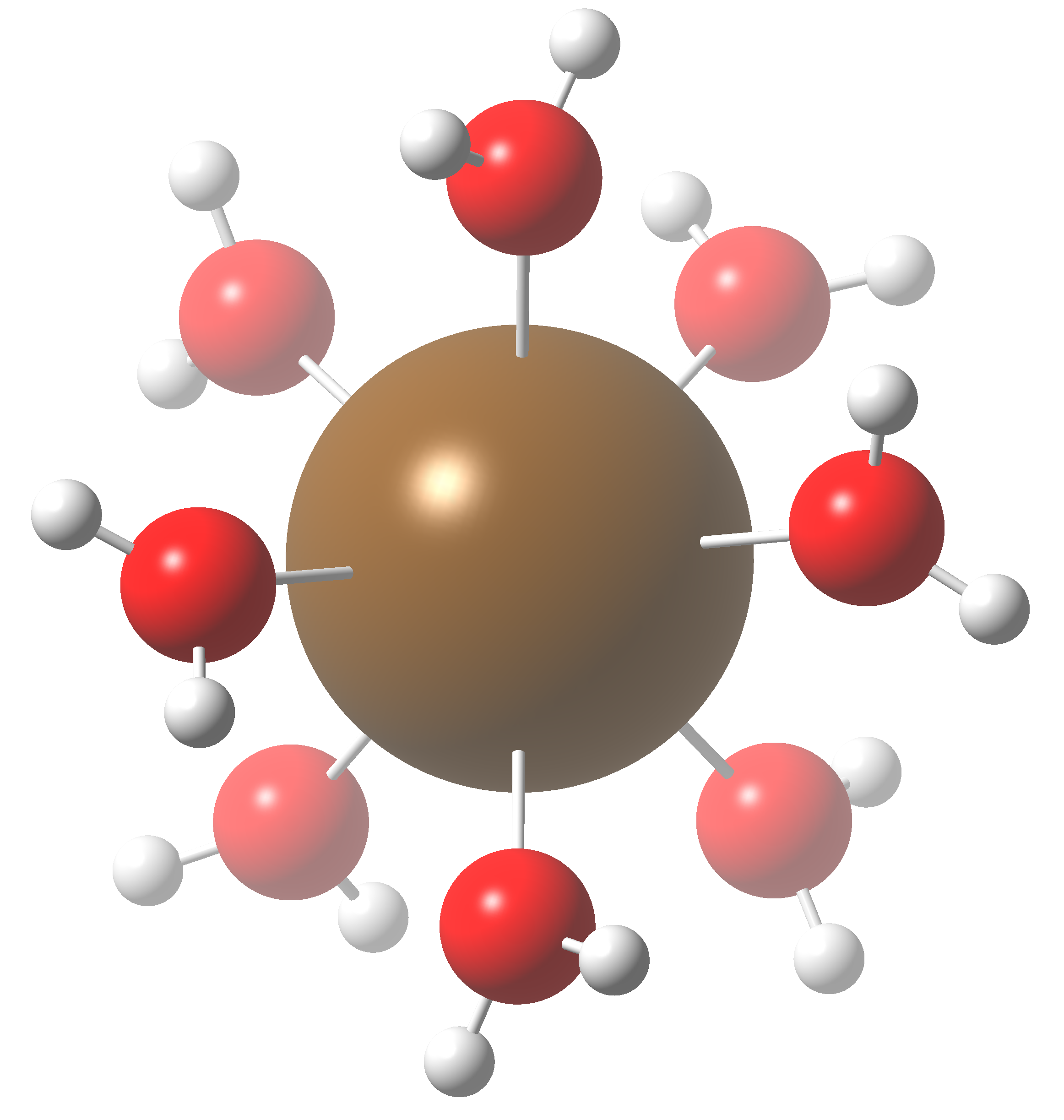}
\includegraphics[width=1.7in]{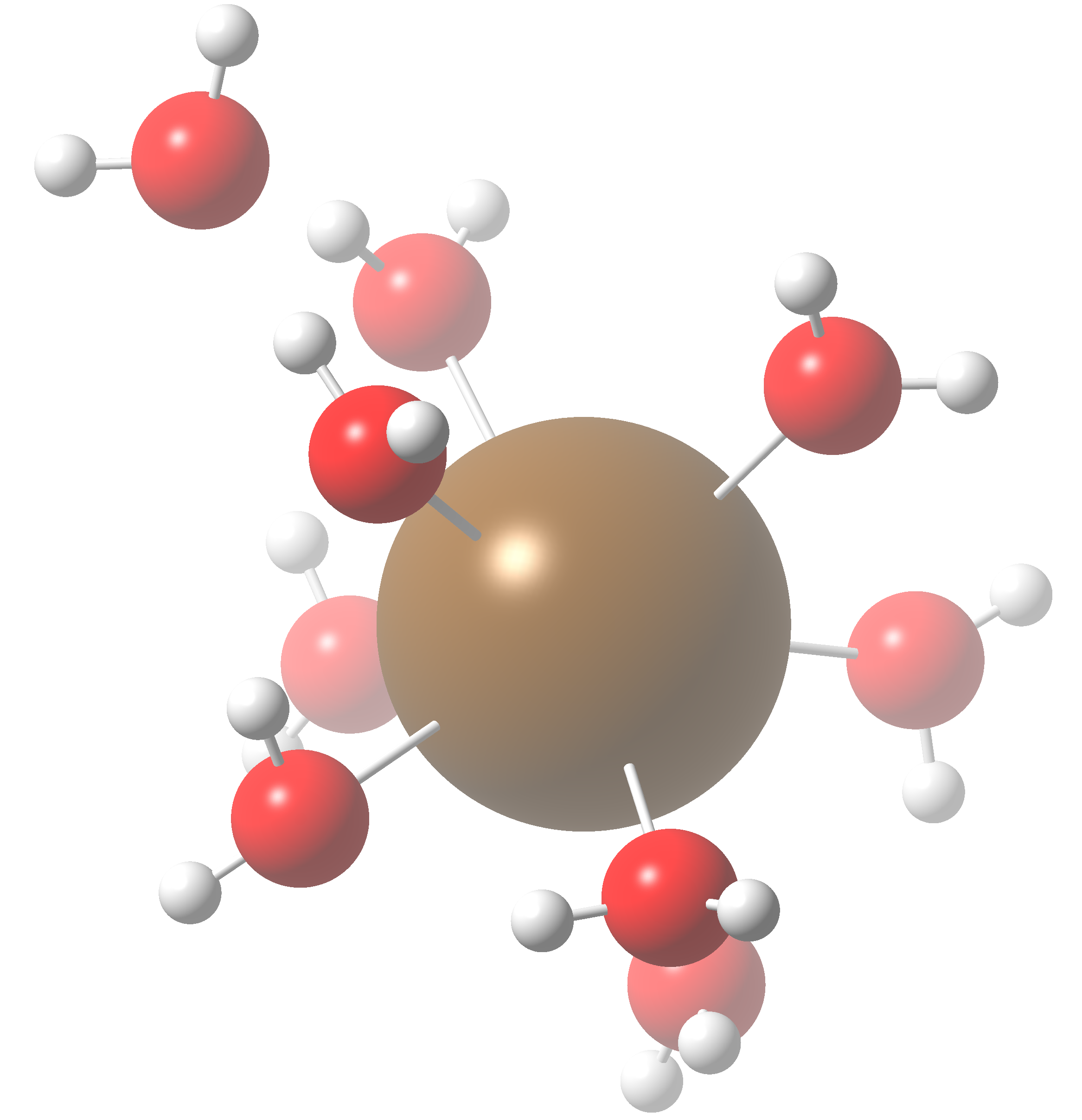}
\caption{Lowest-energy complexes 
with $n$=8 (left) and $n$=9 (right) waters (red oxygens, silver hydrogens)
 around Ba$^{2+}$ ion (brown).  In both structures,
 $n$=1-8 waters occupy the first shell of solvent molecules in a skewed cubic geometry.  
 The 9$^{th}$ water occupies the second shell of solvent, forming a {\em multi-shell} 
 $n$=8+1 complex. Bonds between Ba$^{2+}$ and oxygen atoms represent water molecules closer than the 
 first minimum ($r_\text{min}\approx$ 3.5~{\AA}) in $g(r)$ (FIG.~\ref{fig:gr}).  Nearly identical structures result for the complexes in
 gas phase or surrounded by a polarizable continuum model of liquid water.}
\label{fig:cluster}
\end{center}
\end{figure}

 The location of the ninth water differs
 from an earlier QM/MM study based on
 a Hartree-Fock (HF) description of the quantum mechanical region.\cite{rode:ba}
 That difference is likely due to the neglect of electron correlation effects in the HF level of theory.  
 DFT methods, including the TPSS exchange-correlation functional used in the present work,
 account for electron correlation. Lack of electron
 correlation can result in weaker water-water interactions, thus leading to a smaller water dipole, 
 and thereby an underestimation of water-water repulsion (see, for example, Ref.~76). 
 Weaker water repulsion can presumably manifest as an increase in hydration number, as reported earlier.\cite{rode:ba}
 
The clear division between inner- and outer-shell waters simplifies evaluation of the hydration free energy (Eq.~\ref{eq:1}).  
Still, many choices exist for $\lambda$ and $n$ 
that satisfy the {\it no split shell} rule.  
Previous studies  defined the inner boundary ($\lambda$) to represent either the hydration 
structure of the principle maximum ($r_\text{max}$) \cite{Rogers,Asthagiri:2003kl,Rempe:K,Varma:bj,Asthagiri:2010,Sabo:2013gs} 
or the first minimum ($r_\text{min}$) \cite{Beck:2006wp,Shah:2007dm} in the radial distribution function. 
Thus, we focus our efforts on $\lambda$ values within that range ($r_\text{max}$ $<$ $\lambda \le r_\text{min}$).
We also consider occupancy by $n$=1-8 waters in the first shell and one {\em multi-shell} configuration for comparison, $n$=8+1
(notated $n$=9).  In the latter  configuration (FIG.~\ref{fig:cluster}), 
the first hydration shell is fully occupied and an additional water occupies the second hydration shell.

\begin{figure}
\begin{center}
\includegraphics[width=3.3in]{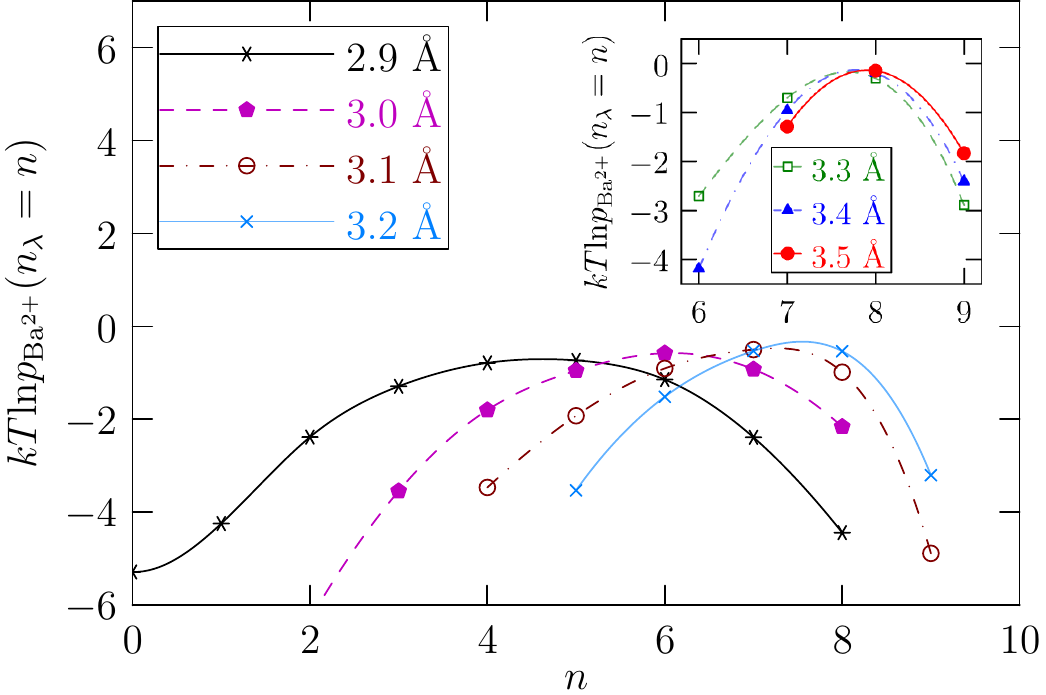}
\caption{Probability distribution for observing $n$ waters within various inner-shell boundaries ($\lambda$, depicted in insets) of Ba$^{2+}$ using AIMD simulation results.
Probabilities given in kcal/mol.  Plots for $\lambda$=3.3-3.5 {\AA} resemble each other and are illustrated separately in inset (top right) for clarity.}
\label{fig:pnx}
\end{center}
\end{figure}

First, we evaluated the population fluctuation contribution, $kT\mathrm{ln}p(n_{\lambda})$, for the chosen 
$\lambda$ values (FIG.~\ref{fig:pnx}).  The full inner-shell population ($n = $ 8) was 
observed for $\lambda$ $>r_\text{max}$, beyond the principle maximum in $g(r)$. 
As expected, the probability $p(n_\lambda)$ for observing small occupancy numbers ($n$)
decreases as the inner-shell  boundary ($\lambda$) increases (FIG.~\ref{fig:pnx}).

\begin{figure}
\begin{center}
\includegraphics[width=3.3in]{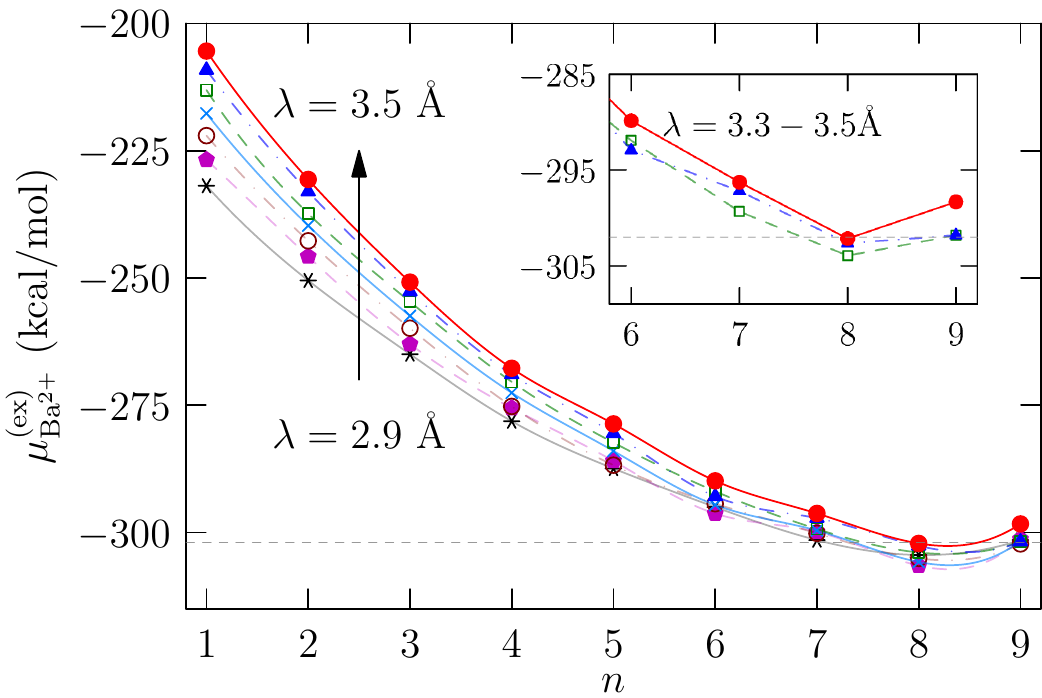}
\caption{Variation in predicted  hydration free energies based on different
 choices for the inner-shell radius ($\lambda$) and coordination number ($n$). The
 horizontal dashed line represents an experimental value of -302.56 kcal/mol.\cite{Marcus:1994ci} The predictions for   
 $n$ = 6-9 converge to within 4 kcal/mol for all $\lambda$ and approach the experimental value with higher $n$ (see inset). 
 The QCT population fluctuation component,
 $kT\mathrm{ln}p_{\mathrm{Ba}^{2+}}({n}_\lambda)$ (FIG.~\ref{fig:pnx}), and an incomplete inner hydration
 shell account for the variations in $\mu^{\mathrm{(ex)}}_{\mathrm{Ba}^{2+}}$
 with $\lambda$ for $n$=1-5.}
\label{fig:hyd_free}
\end{center}
\end{figure}

Then we calculated  $\mu^{\mathrm{(ex)}}_{\mathrm{Ba^{2+}}}$ (Eq. \ref{eq:1})
for the same range of $\lambda$ values using all possible $n$=1-8 inner-shell 
occupancies and the $n$=9 {\em multi-shell} occupancy (FIG.~\ref{fig:hyd_free}).  
  At  smaller $n$=1-5, substantial deviations in $\mu^{(\mathrm{ex})}_{\mathrm{Ba^{2+}}}$ 
with changing $\lambda$ reflect limitations in sampling used to
compute $p(n_\lambda)$,
as well as molecular-scale inaccuracies within the inner hydration shell. The latter makes the larger contribution.
Inaccuracies are anticipated because the outer environment solvates an incomplete inner hydration shell.
For higher $n$=6-9, the inner hydration shell fills to include the waters that saturate the principal maximum
in the radial distribution function.  Also, $n_\lambda \approx \bar{n}_\lambda$ for all inner-shell boundaries ($\lambda$),
permitting the most accurate evaluation of the population term (FIG.~\ref{fig:pnx}).   
For the completely filled inner shell with $n$=8, the mean hydration 
free energy is comparable to the available experimental value of $-302.56$ kcal/mol:\cite{Marcus:1994ci}
$\langle\mu^{(\mathrm{ex})}_{\mathrm{Ba^{2+}}}\rangle$ = $-304.10\pm1.38$ kcal/mol.

\begin{figure}[h]
\begin{center}
\includegraphics[width=3.0in]{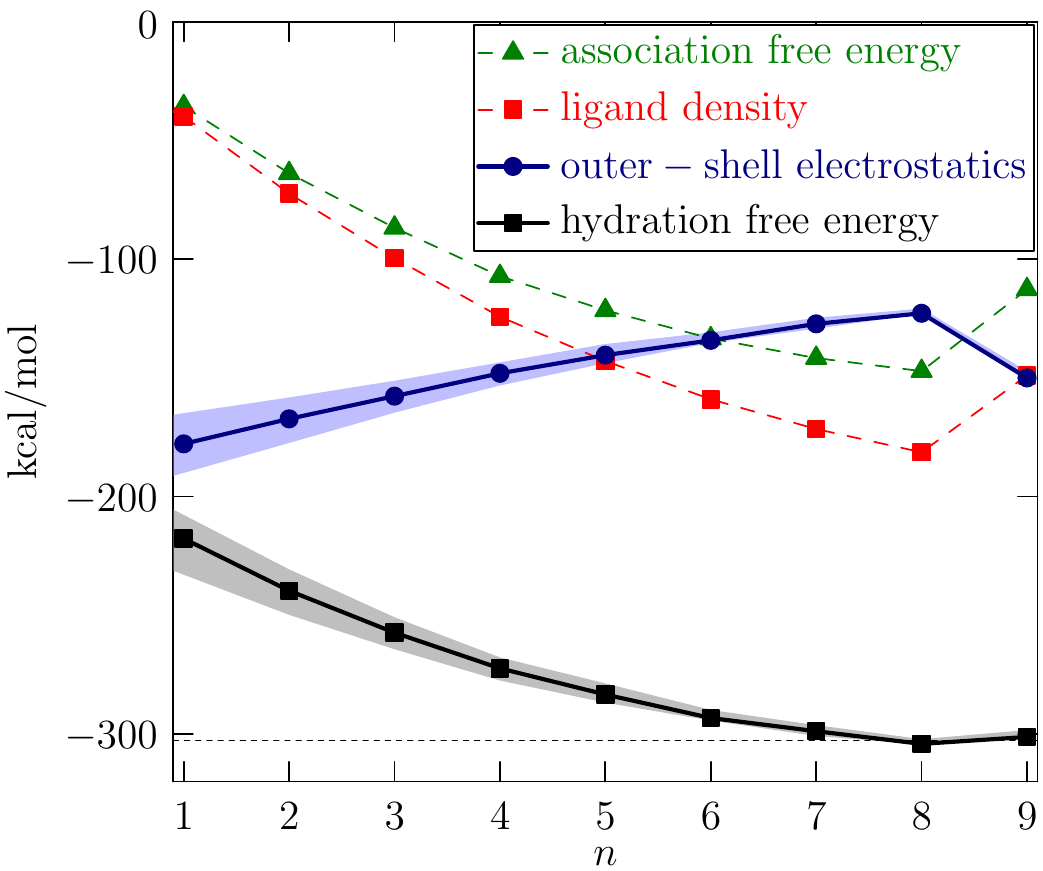}
\caption{Mean free energy contributions to  $\mu^{(\mathrm{ex})}_{\mathrm{Ba^{2+}}}$ (black squares) for
$\lambda$=2.9-3.5~{\AA} (Eq.~\ref{eq:1}):  Ba$^{2+}$-water inner-shell association contributions (green),  
with addition of the ligand density (red), 
and the outer-shell electrostatic contribution that solvates ion-water clusters and desolvates water ligands with a 
dielectric continuum model (blue).  The population fluctuation term ($kT\mathrm{ln}p_{\mathrm{Ba}^{2+}}({n}_\lambda)$) 
was excluded due to its small contribution. Note that all other contributions are substantial, and thus important
for the overall prediction. The outer-shell contribution changes only slightly with $\lambda$ 
(indicated by shading).  The predicted  
 $\mu^{(\mathrm{ex})}_{\mathrm{Ba^{2+}}}$ from 
$n$=6, 7, 8 and the {\em multi-shell} 9 occupancies are 
comparable to the experimental value (horizontal dashed line).\cite{Marcus:1994ci}  The fully occupied
and exclusive inner shell ($n$=8) results in the closest agreement with experiment and the lowest free energy, 
reinforcing the observation that $n$=8 is the dominant structure of the inner hydration shell (FIG.~\ref{fig:pnx}).}
\label{fig:hyd_contri}
\end{center}
\end{figure}

 Individual contributions to $\mu^{(\mathrm{ex})}_{\mathrm{Ba^{2+}}}$ (Eq.~\ref{eq:1}) are shown in 
 FIG. \ref{fig:hyd_contri} for  all $\lambda$ (2.9 - 3.5 \AA).   At larger $\lambda$ values (3.3-3.5 \AA),
the completely filled  inner shell constitutes the most probable occupancy, $\bar{n}_\lambda$ = 8 (FIG.~\ref{fig:pnx}).
Then the mean hydration free energy computed for that subset of  $\lambda$ values
matches the experimental value:\cite{Marcus:1994ci}  $\langle\mu^{(\mathrm{ex})}_{\mathrm{Ba^{2+}}}\rangle$ = $-302.9\pm0.7$ kcal/mol. 
All components of the hydration free energy (Eq.~\ref{eq:1}) make substantial contributions to 
the total except for the population fluctuation term.
Thus, contributions from both local and distant solvent molecules, 
including the contribution from ligand density, are important for the overall prediction.  
Higher occupancy ($n$) results in more favorable inner-shell association contributions because the ion and waters
in gas phase favor clustering.   Higher occupancy also results in less
favorable outer-shell contributions to the free energy because of the penalty associated with desolvating
$n$ waters (Eq.~\ref{eq:1}).  

Those trends in contributions to $\mu^{(\mathrm{ex})}_{\mathrm{Ba^{2+}}}$ with occupancy 
change abruptly for $n$=9 due to the {\em multi-shell} configuration (FIG. \ref{fig:hyd_contri}).  
With addition of the first water to the second hydration shell 
(FIG.~\ref{fig:cluster}), the inner- and outer-shell contributions jump to new values.  The inner-shell association 
free energy is more favorable for $n$=8+1 than for $n$=1, but less favorable than for high occupancy  ($n \geq$6) within the
first hydration shell.  The outer-shell solvation component shows a similar trend, but with reversed sign.  
For even higher occupancy of the second shell, the same trends observed for the first hydration shell are anticipated -- 
more favorable (decreasing) association free energies and less favorable (increasing) outer-shell solvation free energies.  
Adding up the components for the {\em multi-shell} $n$=9 occupancy yields a total hydration free energy slightly less stable than,
but comparable to, the value from the fully occupied and exclusive first hydration shell ($n$=8).
 
 Given that the exchange correlation functional was chosen to reproduce
 experimental inner-shell association free energies,\cite{Peschke1998}  and the population fluctuation contribution is negligible,
 the outer-shell solvation contribution is the term evaluated with the greatest approximation.
 That contribution is modeled by a dielectric continuum treatment of the
 environment interacting with a molecular inner solvation shell.  
 When the inner hydration shell includes the occupancy of the principal maximum
 in $g(r)$ ($n \geq$ 6), predictions of Ba$^{2+}$ hydration free energy show
 reasonable agreement with experiment,\cite{Marcus:1994ci}even with the outer environment treated approximately (FIG.~\ref{fig:hyd_contri}).

 \section{Conclusion}
To summarize, our analysis of the full range of boundary  values (2.9 $\le\lambda\le$ 3.5) indicates that 
once the principal maximum in $g(r)$ is saturated ($n$ $\ge$ 6),
 predictions of hydration free energy are independent of boundary and comparable to the experimental value of $-302.56$ kcal/mol.\cite{Marcus:1994ci} 
That trend persists even with ligand occupancy in the
 second solvation shell (FIG.~\ref{fig:hyd_free}).
But the best predictions result from analysis of the complete and exclusive inner-shell hydration environment, defined for Ba$^{2+}$ at $n$=8 
and obtained for a range of $\lambda$ (3.3-3.5 \AA) for which $n_\lambda$=$\bar{n}_\lambda$=8. 
Then $\langle\mu^{(\mathrm{ex})}_{\mathrm{Ba^{2+}}}\rangle$ is $-302.9\pm0.7$ kcal/mol.

Since Ba$^{2+}$ and K$^+$ are the same size, some similarities in water structure can be expected.   
Comparing with earlier AIMD studies of K$^+$(aq),\cite{Varma:bj} both ions are characterized by the same location 
of the principal maximum in $g(r)$, indicating that waters preferentially
cluster about both ions at the same distance ($r_\text{max}$=2.8 {\AA}).  
The full first hydration shells also extend to 
the same distance ($r_\text{min}$=3.5 {\AA}). 
Nevertheless, the additional charge on Ba$^{2+}$, compared to K$^+$, accounts for notable differences in hydration structure.
Two more waters fill in the principal maximum of Ba$^{2+}$ ($n(r_\text{max})$=6) than K$^+$ ($n(r_\text{max})$=4).  Similarly, two more waters
occupy the first hydration shell of Ba$^{2+}$ ($\left<{{n(r_\text{min})}}\right>$=8) than K$^+$ ($\left<{{n(r_\text{min})}}\right>$=6).
The most significant difference occurs in the inner shell.
Four more waters directly coordinate Ba$^{2+}$ ($n$=8)  than K$^+$ ($n$=4).  
Octa-ligation of Ba$^{2+}\text{(aq)}$, and tetra-ligation of the monovalent ions, agrees with Bernal and Fowler's 
early predictions of local water structure in aqueous solution.\cite{bernal}

Crystal structures of potassium ion channels show Ba$^{2+}$ and K$^+$ directly coordinated by $n$=8 oxygens located 2.8 {\AA} from the ions
in a skewed cubic geometry.\cite{Zhou:2001vo,Jiang:2000,lockless,ye,Jiang:2014}  
In the context of K channels and QCT analysis, recent works have proposed special `quasi-liquid' conditions of high ligand density (similar
to bulk water) and poor ligand solvation environment that stabilize octa-ligation for the permeant K$^+$, but not the smaller
impermeant sodium ion (Na$^+$).\cite{Varma:jacs,Varma:bj,Varma:2011ho} 
Nevertheless, ion binding sites in K channels appear to mimic the local hydration structure of the blocking ion (Ba$^{2+}$), not the
permeant ion (K$^+$).   This result establishes a foundation for further molecular studies of the blocking mechanism and blocking sites
of K channels. 
\section{Acknowledgement}
We thank Lawrence R. Pratt, Sameer Varma, and Dubravko Sabo for helpful discussions. Sandia National Laboratories is a multiprogram laboratory managed and operated by Sandia Corporation, 
a wholly owned subsidiary of Lockheed Martin Corporation, for the U.S. Department of Energy's National 
Nuclear Security Administration under Contract DE-AC04-94AL8500. This work was supported by Sandia's 
LDRD program.

\bibliographystyle{achemso}
\bibliography{baBib}
\clearpage

\end{document}